\documentclass[preprint,prd,showpacs]{revtex4}
\newcommand{\bi}{\begin{itemize}}
\newcommand{\ei}{\end{itemize}}

\newcommand{\be}{\begin{equation}}
\newcommand{\ee}{\end{equation}}
\newcommand{\bdm}{\begin{displaymath}}
\newcommand{\edm}{\end{displaymath}}
\newcommand{\beqa}{\begin{eqnarray}}
\newcommand{\eeqa}{\end{eqnarray}}

\usepackage{epsfig}
\begin{document}
\title{Fermionic states in pure 4D deconstruction}
\author{P. Q. Hung} 
\email[]{pqh@virginia.edu} 
\author{Ngoc-Khanh Tran} 
\email[]{nt6b@virginia.edu}
\affiliation{Department of Physics, University of Virginia \\
382 McCormick Road, Charlottesville, Virginia 22904-4714, USA}
\date{\today}
\begin{abstract}
We study the structure of fermionic mass eigenstates in a pure 
four-dimensional 
deconstruction approach. Unlike the case with the usual higher 
dimensional deconstruction (or latticized extra dimension), here the 
doubling of fermionic degrees of freedom is physical, 
thus there is no need to invoke Wilson terms to eliminate them. 
The fermionic structure is shaped by two key factors, namely the 
boundary conditions on fermions and the ratio of two breaking scales involved.
The singular value decomposition of linear algebra is employed to shed 
light into the phenomenologically crucial role of chiral boundary 
conditions. In this approach, we can explain the ``localization'' 
or ``delocalization'' nature of fermionic zero mode in flavor space 
and obtain analytically all higher modes. The application of boundary 
conditions on fermions to the implementation of CKM quark 
mixing is also found.
\end{abstract}
\pacs{11.25.Mj, 12.15.Ff}
\maketitle
\section{Introduction}
In the past few years, a great amount of renewed interest has been placed 
in the theories with extra spatial dimensions. Intrinsically, 
models in these spaces are non-renormalizable, thus extra dimensional theory 
can be seen generally as an effective one, being valid only below some 
cut-off scale. The important issue of cut-off independence or dependence 
in such models, though could be argued in specific contexts, remains 
rather obscured. With the advent of the dimensional deconstruction (DD) 
concept 
\cite{ACG,HPW}, there exists the possibility to construct extra 
dimensional (ED) scenarios at lower 
energy scale effectively from a renormalizable four-dimensional (4D) theory. 
The chiral nature of fermions at low energy as dictated by the standard 
model (SM) then requires the imposition of some type of chiral boundary 
conditions (CBC) on the fermions. The simplest CBC mimics the Dirichlet 
and Neumann boundary conditions used in the orbifold compactification 
of extra dimension (see e.g. \cite{HPW}). In this work, making use of the 
singular value decomposition of linear algebra, we systematically  
identify and analyze the role of general classes of CBCs  
on the fermion mass-eigenstates after the deconstruction. 
It turns out in particular that the chiral zero-mode wave function 
possesses a non-trivial, localized or delocalized, distribution in the 
flavor space \cite{SS,AKMY}. 

The pattern of localization is found to depend exclusively on two  
factors. The first of them is the ratio of the two underlying breaking 
scales present in the general deconstruction scenario, namely the link 
field vacuum expectation value (VEV) and the fermion bare mass (which can 
also be seen dynamically as the VEV of a gauge-singlet scalar field). The 
other factor is the chiral boundary condition being employed. 

To study the effect coming solely from the 
ratio of these VEVs on the localization, we first present an 
analytical derivation of fermion  
zero-mode wave function under the Dirichlet-Neumann CBC. In the flavor space, 
the link field VEV sets the ``dimension'' where fermions of all modes can live 
on, while the fermion bare mass sets the size of the ``domain wall'' 
wherein zero-modes are confined. The relative magnitude of these two 
scales then obviously has a crucial role in shaping the zero-mode 
localization. And for an appropriate choice 
of this ratio, the whole phenomenological applicability of fat-brane 
models is carried over to the dimensional deconstruction scenario due 
the similar localization behavior. 

To see the fermion wave function's dependence on the 
boundary conditions, we next work with more general CBCs {\em a priori}. 
Here, the singular value decomposition (SVD) theorem proves to be a 
powerful tool in the systematic identification of CBC, with which the 
spectrum and wave functions of all modes can be exactly solved. In 
this sense, selecting boundary conditions also means modeling the outcome 
of the Kaluza-Klein (KK) spectrum. Furthermore, many related 
quantities, those are important in model building such as 
phase and wave function overlaps, can now be encoded concisely in the 
boundary conditions. 

In the other perspective, it may be also tempting to geometrically embed 
the deconstruction group into 
a latticized extra dimension as has been done in \cite{CHPW,HL}. However, 
the well-known fermion flavor doubling problem in lattice models, or else 
the need of 
a faithful lattice representation of the continuum, requires some remedy such 
as the incorporation of Wilson terms into the Lagrangian. These terms however  
eliminate half of fermionic chiral degrees of freedom (see Appendix). It is 
for this reason that, in the literature, as far as the fermion sector 
is concerned, 
one usually starts with Weyl rather than Dirac spinors \cite{HPW,SS,AKMY}. 
In this paper, motivated by the ability of dimensional deconstruction in  
providing the UV completion for 
ED models, we perform the deconstruction purely within the 4D framework 
(4D deconstruction), and will not build an actual latticized extra 
dimension to host the deconstruction product group 
\footnote{However, we will still use the ED picture to enhance the 
visualization effect of physical explanation wherever it is appropriate.}. 
This is equivalent to 
beginning with a set of Dirac spinors. It is shown that as long as the zero 
modes are concerned, the 4D deconstruction is not more complicated than the 
latticized ED model. Along this line of reasoning, it is worthwhile to 
note that in a 4D 
deconstruction context, fermions do not actually live in any exotic orbifold,
and the concept of ``boundary condition'' mentioned above in reality can be 
perceived rather as a defect in some deconstruction 
group representation. That is why  fermions of different flavors do not 
necessarily obey the same boundary conditions. And this fact can be used to 
model the quark families hierarchical mixings observed in standard 
model.  

This work is structured as follows. In Section II we derive and analyze  
the localization patterns of zero mode fermion after the 4D deconstruction. 
The chiral nature of zero mode is also visualized in the analog 
coupled oscillator system. We recall the SVD theorem in Section IIIA
and employ it to clarify the crucial role of CBC in 
the structuring of fermion mass eigenbasis in Section IIIB. 
Then we illustrate how complex CKM mixings and mass hierarchy 
of SM quarks could arrive from the 
appropriate selection of boundary condition in Section IIIC. 
We summarize the main 
results in Section IV. Finally, for the purpose of comparison, in the 
Appendix we recapitulate the essence of latticized ED scenario.
\section{Localization of chiral zero modes in flavor space}
In the general dimensional deconstruction approach, fermions associated 
with different gauge groups  
are coupled to one another via scalar link fields through Yukawa interaction 
terms. By giving appropriate vacuum expectation value (VEV) to these link 
fields one can completely restructure the fermion mass spectrum and keep 
only its lightest modes to be the phenomenologically relevant ones at low 
energy. Such zero mode exposes some very interesting ``localization'' 
behavior in the group index space (also referred to as flavor space 
throughout), parallel to that of the 
split fermion scenario in extra dimension. Thus the pure 4D deconstruction 
mechanism can have vast applicabilities to phenomenology although no 
actual extra dimensions have been invoked.
\subsection{Localization formation}
The general setup of dimensional deconstruction contains a gauge group 
product $\prod_{n=1}^{N} G_n$ and $N$ fermionic flavors $\psi_n$ 
transforming fundamentally under the corresponding group $G_n$ in the usual 
4D space time. 
There are also $N-1$ scalars $\phi_{n,n+1}$ living in the 
fundamental and anti-fundamental representation of  
$G_n$ and $G_{n+1}$ respectively. Hence each scalar couples to two 
``neighboring'' groups and can be also referred to as a link field. For 
simplicity, we also assume the universal fermion-scalar-fermion couplings, 
which can be realized by imposing a permutation symmetry concerning 
the group index $n$. The gauge-invariant Lagrangian 
describing this ``open moose'' set-up reads as follows
\be
\label{L}
{\cal{L}}=\sum_{n=1}^{N} \bar{\psi}_n i{\not\!\!D}_n \psi_n +
\sum_{n=1}^{N-1} \bar{\psi}_n \phi_{n,n+1} \psi_{n+1} - 
\phi_S \sum_{n=1}^{N} \bar{\psi}_n   \psi_n 
\ee
where $\phi_S$ denotes a gauge-singlet scalar and ${\not\!\!D}_n$ the 
covariant derivative. In respect to the above permutation symmetry, 
the following spontaneous symmetry breakings 
\be
\phi_S  \rightarrow \langle \phi_S \rangle = M; \;\;\;\;\;\;\; 
\phi_{n,n+1} \rightarrow \langle \phi_{n,n+1} \rangle = V  \;\;\;\; 
(\forall n) 
\ee
give rise to a fermion mass structure. The gauge singlet vacuum 
expectation value $M$ is also referred to as fermion bare mass hereafter. 
In the chiral basis
\be
\label{Ncolumn}
\{\psi_{L,R}^N\} \equiv \{ \psi_{1L,R},\ldots,\psi_{NL,R}\}^T; 
\;\;\;\;\;\;\;\;\;\;
\{\bar{\psi}_{L,R}^N\}\equiv\{\bar{\psi}_{1L,R},\ldots,\bar{\psi}_{NL,R}\} 
\ee
the mass term can be written as
\be
\label{mass}
\{\bar{\psi}_L^N\} \;[{\cal{M}}] \; \{\psi_R^N\} 
+ \{\bar{\psi}_R^N\} \; [{\cal{M}}] \; \{\psi_L^N\}
\ee
with $[{\cal{M}}]$ being the $N\times N$-dimension mass matrix
\be
\label{MLR}
[{\cal{M}}]_{N\times N}= [{\cal{M}}]^{\dagger}_{N\times N}=
\left( \begin{array}{ccccccc}
M&-V&0& & & &  \\
-V^{*}&M&-V& & & &  \\
 0&-V^{*}&M  & & & &  \\
&&&\ddots&&& \\
&&&&M&-V&0 \\
&&&&-V^{*} &M&-V \\
&&&&0&-V^{*}& M
\end{array} \right) 
\ee
Each link field transforms non-trivially under two groups and V assumes  
complex value in general 
\footnote{One can always construct the link field potential and choose its 
parameters, so that a complex VEV is produced \cite{BLS}}. 
In contrast, the gauge singlet vacuum expectation value M is 
always real as required by the hermiticity of the Lagrangian (\ref{L}). 
It is obvious from Eq. (\ref{MLR}) that this setup has all chiral fermion  
pairs degenerate in mass and thus describes a vector-like model. As in 
theories with EDs, in order to have a chiral model at low energy limit we need 
some chiral boundary conditions (CBC). In the context of dimensional 
deconstruction, these ``CBCs'' can be realized by adopting the following  
gauge-invariant asymmetric choice of fermion content in the chiral 
representation space 
\footnote{Other CBCs can be transformed into (\ref{CBC}) by a permutation 
or an unitary rotation (see Section III).}
\be
\label{CBC}
\psi_{1R}=\psi_{NR}=0; \;\;\;\;\;\;\;\;\;\; 
\phi_{k-1,k} \psi_{k,L} = V \psi_{k-1,L}   
\ee
In the truly extra dimensional set-up, the boundary conditions being 
asymmetric with respect to fermionic chiral components 
(i.e. chiral boundary conditions) are shown to be compatible with the 
variational principle \cite{CGHST} in e.g. Higgsless models \cite{CGPT}. In 
4D deconstruction, the CBCs (\ref{CBC}) literally are the ``defects'' in  
fermion representation at some particular sites. At least in the  
large $N$ (i.e. continuum) limit, these CBCs too 
might follow from the application of variational principle on some 
appropriately constructed action. This approach to the 4D deconstruction 
CBCs is currently under our investigation.   

The first ``boundary condition'' of (\ref{CBC}) is simply the statement that 
$\psi_{1L}$ and $\psi_{NL}$ do not have matching right-handed counterparts. 
Notice also that the above boundary conditions break our permutation 
symmetry, a feature which is very analogous to the breaking of translational 
symmetry in compact extra dimensions by orbifold boundary conditions.  
When the link fields assume VEV $\langle \phi \rangle = V$ and $k=2$ 
(or $N$), Eq. (\ref{CBC}) mimics the Dirichlet and Neumann boundary 
conditions 
\cite{HPW}. However, in pure 4D deconstruction, there is no latticized 
ED to host the group chain. That is why there is no need to identify $k$ to 
the ``end-points'' $k=2$ (or $N$). 
Rather one can choose k to be any integer $\in [2,N]$. We can also write  
(\ref{CBC}) in a systematic matrix notation
\beqa
\label{CBCL}
\left( \begin{array}{c}
\psi_{1L}\\
\\
:\\
:\\
\\
\psi_{NL} 
\end{array} \right)=
\{\psi_L^N\} &=& [{\cal{B}}_L]_{N\times (N-1)}\{\psi_L^{N-1}\}=
\left( \begin{array}{ccccc}
1&&& &    \\
&:&& &    \\
&&1&& \\
&&1&& \\
&&&:& \\
&&&&1
\end{array} \right)
\left( \begin{array}{c}
\psi_{1L}\\
:\\
\psi_{k-1L}\\
\psi_{k+1L}\\
:\\
\psi_{NL} 
\end{array} \right) ; 
\\
\label{CBCR}
\left( \begin{array}{c}
\psi_{1R}\\
\\
:\\
\\
\psi_{NR} 
\end{array}\right)
=\{\psi_R^N\} &=& [{\cal{B}}_R]_{N\times (N-2)}\{\psi_R^{N-2}\}=
\left( \begin{array}{cccccc}
0&&& & &0   \\
1&&& & &   \\
&&:&&& \\
&&&&&1 \\
0&&&&&0
\end{array} \right)
\left( \begin{array}{c}
\psi_{2R}\\
\\
:\\
\\
\psi_{N-1R} 
\end{array}\right)
\eeqa
where $\{\psi_L^{N-1}\}$ and $\{\psi_R^{N-2}\}$ as defined in (\ref{CBCL}), 
(\ref{CBCR}) denote truly independent $N-1$ left-handed and $N-2$ 
right-handed fermionic degrees of freedom. Non-square matrices 
$[{\cal{B}}_L]$ and $[{\cal{B}}_R]$ precisely encode the chiral boundary 
conditions (\ref{CBC}). These CBCs aim to break the L-R symmetry 
in the mass term (\ref{mass}), which now becomes
\be
\label{Mmass} 
\{\bar{\psi}_L^{N-1}\} \;[{\cal{M}}_{LR}] \; 
\{\psi_R^{N-2}\} +
\{\bar{\psi}_R^{N-2}\} \;[{\cal{M}}_{RL}] \; 
\{\psi_L^{N-1}\}
\ee
with chiral mass matrices
\be
\label{mLR}
[{\cal{M}}_{LR}]_{(N-1)\times (N-2)} \equiv 
[{\cal{B}}_L]^{\dagger}[{\cal{M}}][{\cal{B}}_R];
\;\;\;\;\;\;\;
[{\cal{M}}_{RL}]_{(N-2)\times (N-1)} \equiv 
[{\cal{B}}_R]^{\dagger}[{\cal{M}}][{\cal{B}}_L]
\ee
By coupling the chiral Dirac equations 
\be
\label{DiracEq}
i\not{\partial}\{\psi_L^{N-1}\} - [{\cal{M}}_{LR}]\{\psi_R^{N-2}\} = 0; 
\;\;\;\;\;\;\;\;\;
i\not{\partial}\{\psi_R^{N-2}\} - [{\cal{M}}_{RL}]\{\psi_L^{N-1}\} = 0
\ee
we see that the squared-mass matrix is 
$[{\cal{M}}^2_{L}]_{(N-1)\times (N-1)} 
\equiv [{\cal{M}}_{LR}][{\cal{M}}_{RL}] $ for the 
left-handed components $\{\psi_L\}$ and 
$[{\cal{M}}^2_{R}]_{(N-2)\times (N-2)} 
\equiv [{\cal{M}}_{RL}][{\cal{M}}_{LR}] $ for the 
right-handed $\{\psi_R\}$. In Section III we will show that under any 
CBCs of the type (\ref{CBC}), $[{\cal{M}}^2_{L}]$ always possesses zero 
eigenvalue, assuring the identification of its corresponding eigenstate with  
SM left-handed fermions. For now we just present the 
analytical derivation of this chiral massless eigenstate for any 
values of $M$ and $V$. 

First we note that because $[{\cal{M}}_{LR}]$ has dimension 
$(N-1)\times (N-2)$, its left inverse matrix $[{\cal{M}}^{-1}_{LR}]$ 
exists (i.e. $[{\cal{M}}^{-1}_{LR}][{\cal{M}}_{LR}]={\bf{1}}$). Therefore 
the solution of zero eigen-problem of 
$[{\cal{M}}^2_{L}]$ is identical to  that of $[{\cal{M}}_{RL}]$, i.e.
\be
\label{Eqxn}
[{\cal{M}}^2_{L}] \{x_n^{N-1}\} = 0 \Leftrightarrow
[{\cal{M}}_{RL}]  \{x_n^{N-1}\} = 0 
\ee
where we have $N-1$ variables 
$\{x_n^{N-1}\}\equiv \{ x_1,\ldots,x_{k-1},x_{k+1},\ldots,x_N \}^T $ 
in the notation of (\ref{CBCL}). Using the definition (\ref{mLR}) 
of $[{\cal{M}}_{RL}]$, we can write (\ref{Eqxn}) explicitly as
\be
\label{eqxn}
\rho^* x_{n-1} - x_n + \rho\; x_{n+1} = 0 
\;\;\;\;\;\;\; (n=1,\ldots,N)
\ee
with the newly introduced variable $x_k$ being defined as $x_k\equiv x_{k-1}$ 
and  
\be
\label{rho}
\rho \equiv \frac{V}{M}\equiv \frac{|V|e^{i\theta} }{M}
\ee 
Since all link fields have been uniformly broken, the coefficients of 
(\ref{eqxn}) are all independent of index $n$. This in turn suggests a 
solution of the form $x_n= a \exp{(bn)}$. We obtain
\be
\label{solxn}
x_n=
\left\{ \begin{array}{lll}
{\cal{C}}\;e^{-in\theta}
[\sinh{(k-n)\alpha}-e^{i\theta}\sinh{(k-1-n)\alpha}] \;\;\;\;\;\;\;
& \mbox{if $|\rho| < 1/2$} 
& (\cosh{\alpha}\equiv 1/2|\rho|) \\
1/\sqrt{N-1} \;\;\;\;\;\;\;
& \mbox{if $|\rho| = 1/2$} & \\
{\cal{C}}\;e^{-in\theta}
[\sin{(k-n)\alpha}-e^{i\theta}\sin{(k-1-n)\alpha}] \;\;\;\;\;\;\;
& \mbox{if $|\rho| > 1/2$} & (\cos{\alpha}\equiv 1/2|\rho|)
\end{array} \right.
\ee
with the normalization factor ${\cal{C}}$ satisfying 
\be
\label{normalization}
\sum_{n \neq k}^{N} |x_n|^2 =1
\ee
Physically, $\{x_n\}$ is interpreted 
as the weight of the zero mass 
eigenstate $\tilde{\psi}_0$ distributed in the flavor space $\{\psi_L\}$
\be
\label{WF}
\tilde{\psi}_0 =\{x_n^{N-1}\}^{\dagger}\{\psi_{L}^{N-1}\}  
=\sum_{n\neq k}^{N} x_n^* \psi_{nL}
\ee
Because of (\ref{normalization}) and (\ref{WF}), $\{x_n\}$ can be referred 
to as ``wave function'' of fermion zero mode in the group index space. 
The explicit solution (\ref{solxn}) reveals some very interesting properties 
of the massless eigenstate. First, this state is genuinely complex 
as long as V is. Second, (\ref{solxn}) may have a localized or oscillatory 
behavior in the group index space depending on the value of underlying 
parameters (see Section IIB). Both of these properties have 
important applications among others to the characterization of CP violation 
and fermion mass hierarchy in the Standard Model. We now will explore  
quantitatively the behavior of this localization mechanism in the dimensional 
deconstruction scenario. 
\subsection{Pattern of localization}
The chiral zero mode wave function (\ref{solxn}) has a clear dependence on 
the ratio $\frac{V}{M}$ of link field and gauge-singlet scalar VEVs. 
For a pure view on the nature 
of localization, we can just assume for the moment that V is real 
(i.e. $\theta =0$ in (\ref{rho})). In this case (\ref{solxn}) looks 
particularly simple
\be
\label{Rsolxn}
x_n=
\left\{ \begin{array}{lll}
{\cal{C}}\;\cosh{(k-n-1/2)\alpha} \;\;\;\;\;\;\;
& \mbox{if $\rho < 1/2$} 
& (\cosh{\alpha}\equiv 1/2\rho) \\
1/\sqrt{N-1} \;\;\;\;\;\;\;
& \mbox{if $\rho = 1/2$} & \\
{\cal{C}}\;\cos{(k-n-1/2)\alpha} \;\;\;\;\;\;\;
& \mbox{if $\rho > 1/2$} & (\cos{\alpha}\equiv 1/2\rho)
\end{array} \right.
\ee

\begin{figure}
        \begin{center}    
        \epsfig{figure=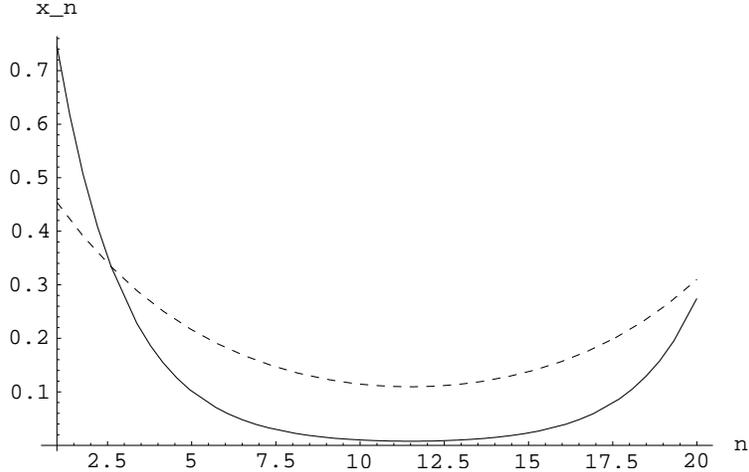,width=0.6\textwidth}
        \end{center}
        \vspace{0cm}
        \caption{Normalized wave function profile (\ref{Rsolxn}) with k=12   
in flavor space. $\rho=0.44$ (or $\alpha =0.52$) for continuous line 
and $\rho=0.49$ (or $\alpha =0.20$) for dashed line. N=20 in both cases. 
The wave function with smaller $\rho$ is more localized.}
        \label{Fig1}
        \end{figure}

\begin{figure}
        \begin{center}    
        \epsfig{figure=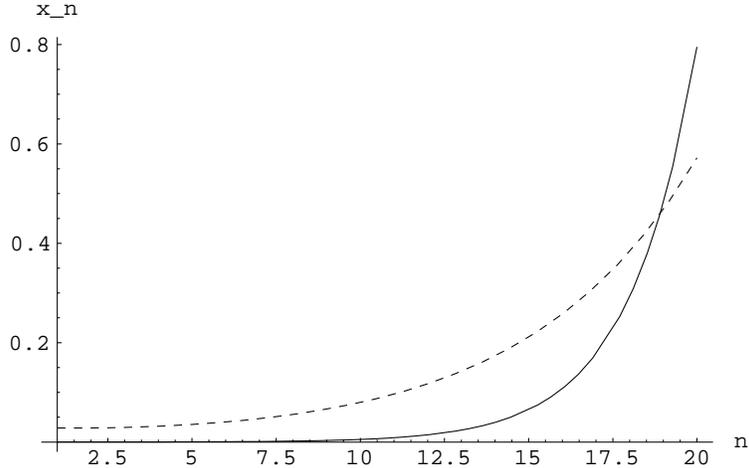,width=0.6\textwidth}
        \end{center}
        \vspace{0cm}
        \caption{Normalized wave function profile (\ref{Rsolxn}) with k=2  
in flavor space. $\rho=0.44$ (or $\alpha =0.52$) for continuous line 
and $\rho=0.49$ (or $\alpha =0.20$) for dashed line. N=20 in both cases. 
The wave function with smaller $\rho$ is more localized.}
        \label{Fig2}
        \end{figure}

\begin{figure}
        \begin{center}    
        \epsfig{figure=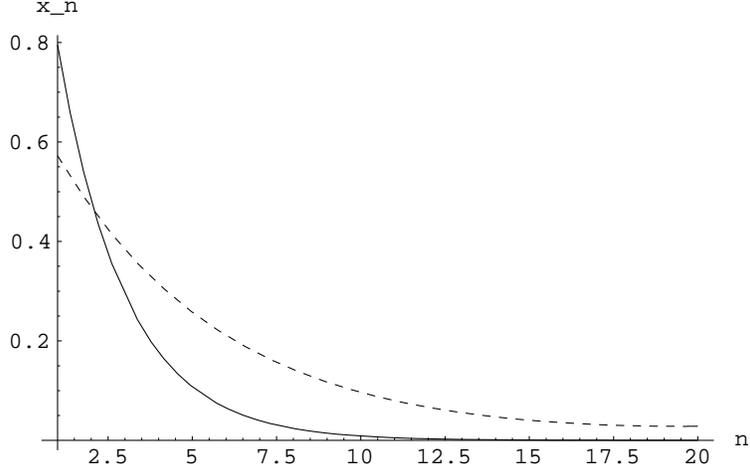,width=0.6\textwidth}
        \end{center}
        \vspace{0cm}
        \caption{Normalized wave function profile (\ref{Rsolxn}) with k=20  
in flavor space. $\rho=0.44$ (or $\alpha =0.52$) for continuous line 
and $\rho=0.49$ (or $\alpha =0.20$) for dashed line. N=20 in both cases. 
The wave function with smaller $\rho$ is more localized.}
        \label{Fig3}
        \end{figure}

For $\rho \equiv V/M < 1/2$, zero mode wave function has a localized 
profile described by a $\cosh$ function (Fig. \ref{Fig1}). 
This can be explained both 
physically and mathematically as follows. As $\rho < 1/2$, the linkage 
symmetry breaking induced mass (proportional to $V$) of fermions 
is essentially smaller 
than their bare mass $M$. In an intuitive approach inspired by the  
ED perspective, $L_V \equiv \frac{1}{V}$ sets the size 
of the ``extra dimension'' (see Eq. (\ref{Vm}) below) and 
$L_M \equiv \frac{1}{M}$ sets the size of the ``domain wall'', to which chiral 
fermion is trapped in the discretized version of Jackiw-Rebbi localization 
mechanism \cite{CHW,JR}.   
\footnote{As the bare mass $M$ is dynamically generated through the 
breaking of a gauge-singlet (background) scalar $\phi_S$ (Eq. (\ref{L})), 
the analogy with the original Jackiw-Rebbi mechanism \cite{JR} is clear.} 
Clearly, localization makes sense only when 
$L_M \ll L_V$ (i.e. $\rho \ll 1$). 
In a more quantitative approach, as $V<\frac{M}{2}$, the gauge 
eigenstates $\{\psi^n_{L,R}\}$ well resemble the mass eigenstates 
$\{ {\tilde{\psi}}^n_{L,R} \}$.
And when $k\neq 2$ (and $N$), the CBC 
(\ref{CBC}) on right-handed fields $\psi_{1R}=\psi_{NR}=0$ implies that 
their opposite (unpaired) chiral partners $\psi_{1L}$ and $\psi_{NL}$ remain 
approximately massless chiral states. That is why in Fig. \ref{Fig1} 
we see that 
the exact chiral zero mode indeed is dominated by $\psi_{1L}$ 
and $\psi_{NL}$. In the special cases where $k=2$ (or $N$), the CBC 
(\ref{CBC}) on left-handed fields eliminates $\psi_{1L}$ (or $\psi_{NL}$) 
from the set of truly independent fermionic degrees of freedom. Then the 
exact chiral zero mode is dominated only by $\psi_{NL}$ (or $\psi_{1L}$), i.e. 
localization is around ``fixed-point'' n=N in Fig. \ref{Fig2} 
(or n=1 in Fig. \ref{Fig3}). 

In all cases, the localized wave functions are exponentially 
suppressed into the middle values of $n$. Let us also remind ourselves  
that $k=2$ (or $N$) corresponds to the familiar Dirichlet-Neumann boundary 
conditions in literature \cite{HPW}. 

\begin{figure}
        \begin{center}    
        \epsfig{figure=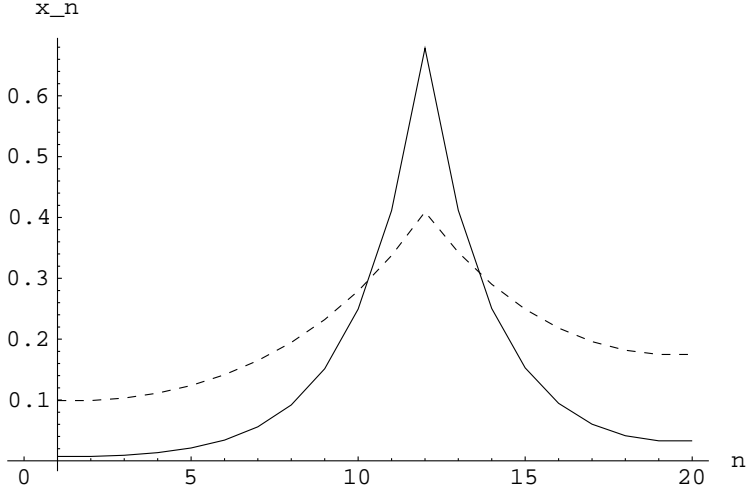,width=0.6\textwidth}
        \end{center}
        \vspace{0cm}
        \caption{Normalized wave function profile (\ref{ksolxn}) with k=12   
in flavor space. $\rho=0.44$ (or $\alpha =0.52$) for continuous line 
and $\rho=0.49$ (or $\alpha =0.20$) for dashed line. N=20 in both cases. 
The wave function with smaller $\rho$ is more localized.}
        \label{Fig4}
        \end{figure}

Mathematically, from Eq. (\ref{eqxn}), 
$\rho <1/2$ requires that $x_{n-1} + x_{n+1} > 2 x_n$, i.e. wave function 
profile should be concave. Together with the boundary condition 
$x_k=x_{k-1}$, this implies $x_1>\ldots > x_{k-1} = x_k < \ldots <x_N$, 
so that wave function is ``pushed'' away from site n=k. 
Finally, in practical application, one may want to localize the chiral 
zero mode around some arbitrary site $n=k$. This can be done very 
effectively by simply combining the two localization patterns presented 
above, i.e. by the imposition of the following boundary conditions
\beqa
\label{CBC1}
&&\psi_{1R}=\psi_{kR}=0 \;\;\;\;\;\;\;
{\phi}_{1,2} \psi_{2L} = V \psi_{1L} \\
\label{CBC2}
&&\psi_{kR}=\psi_{NR}=0 \;\;\;\;\;\;\;
{\phi}_{N-1,N} \psi_{NL} =V \psi_{N-1L}
\eeqa 
In this case, (\ref{CBC1}) makes $x_1\ll x_k$, while (\ref{CBC2}) makes 
$x_k\gg x_N$, so that the localization around $n=k$ is realized.  
Indeed, for these combined BCs, the detailed computation yields the 
following  wave function distribution in the group index space 
(Fig. \ref{Fig4}):
\be
\label{ksolxn}
x_n=
\left\{ \begin{array}{ll}
{\cal{C}}\;\cosh{[(N-k-1/2)\alpha]} \cosh{[(n-3/2)\alpha]} \;\;\;\;\;\;\;
& \mbox{for $1 \leq n \leq k$}  \\
{\cal{C}}\;\cosh{[(N-n-1/2)\alpha]} \cosh{[(k-3/2)\alpha]} \;\;\;\;\;\;\;
& \mbox{for $k \leq n \leq N$}
\end{array} \right.
\ee
where $\cosh{\alpha}\equiv 1/2\rho$ and ${\cal{C}}$ is the normalization 
constant determined by $\sum_{n=2}^{N-1} |x_n|^2 =1$. The overlap of 
two wave functions localized around $n=k_1$ and $n=k_2$ can be exactly 
computed from the analytical expression (\ref{ksolxn}). In the leading 
order it is  
\be
\sum_{n=1}^N x^{(1)}_n x^{(2)}_n \sim e^{-|k_1 \alpha_1- k_2 \alpha_2|}
\ee
Evidently, the overlap is exponentially suppressed by the separation between 
localization centers. It also depends on the localization pattern 
of the underlying  wave 
functions characterized by parameters $\alpha_1$, $\alpha_2$. This property 
is a strong reminiscence of the spit fermion model in the extra dimension. 

\begin{figure}
        \begin{center}    
        \epsfig{figure=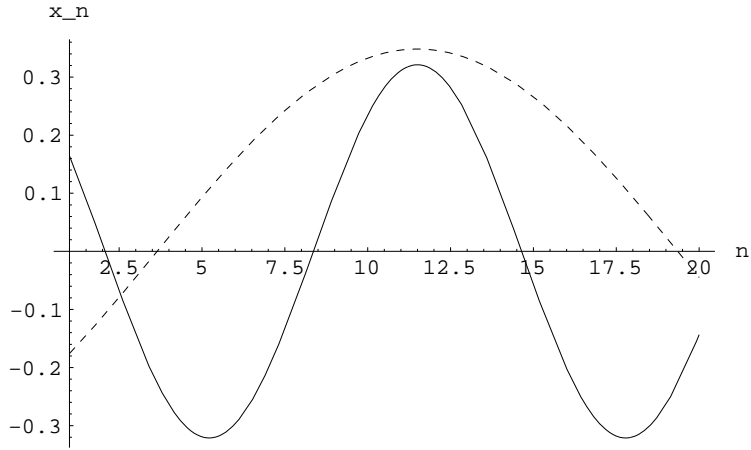,width=0.6\textwidth}
        \end{center}
        \vspace{0cm}
        \caption{Normalized wave function profile (\ref{Rsolxn}) with k=12   
in flavor space. $\rho=0.57$ (or $\alpha =0.50$) for continuous line 
and $\rho=0.51$ (or $\alpha =0.20$) for dashed line. N=20 in both cases. 
The wave function with larger $\rho$ exposes a truly oscillatory pattern.}
        \label{Fig5}
        \end{figure}

For $\rho \equiv V/M > 1/2$, the symmetry breaking effect prevails  
over the bare mass. Even with CBC (\ref{CBC}), 
gauge eigenstates $\psi_{1L}$ and $\psi_{NL}$ no longer necessarily dominate  
the massless eigenstates. The massless mode oscillates like a 
trigonometric function (\ref{Rsolxn}) in the group index space. However, 
in this case the number of groups $N$ also plays an important role. One can 
have a truly oscillatory wave function (Fig. \ref{Fig5}) only if $N$ 
is larger than 
the oscillation period ($\approx 2\pi/\alpha$). Else if $N< \pi/\alpha$, 
the CBC (\ref{CBC}) produces a weakly (quasi) localized zero mode wave 
function around $n=k$ (Fig. \ref{Fig5}). 

The analysis of the case $\rho= 1/2$ can offer a physical insight on the 
single chirality nature of the  massless mode and will be presented next. 
\subsection{Perfect delocalization}
We first note that when $|\rho|= 1/2$, the system (\ref{eqxn}) has a  
solution only if $\theta =0$ or $2\pi$. This implies that for 
$|\rho|=1/2$, CBC (\ref{CBC}) is compatible only with real $V$. In 
this case the zero mode is real and independent of index $n$, i.e. the wave 
function is flat (or perfect delocalization) in the group index 
space (\ref{solxn}). 

We leave a rigorous analysis of the chiral boundary conditions in 
deconstruction mechanism for Section III. For now, however, we are 
content with presenting a visualized physical picture on the chiral nature of 
zero modes by drawing the connection between dimensional deconstruction 
and the classical spring-ball chain system. The massless left and 
right-handed mode, 
if they exist, must be the solution of following equations (in the notation 
of Eq. (\ref{Eqxn}))
\be
[{\cal{M}}^2_{L}] \{x_n^{N-1}\} = 0; \;\;\;\;\;\;\;\;
[{\cal{M}}^2_{R}] \{y_n^{N-2}\} = 0
\ee
where $x$ and $y$ respectively denote left and right-handed solution. The 
analysis below holds for any value of $k$ in Eq. (\ref{CBC}), but to 
keep the presentation 
simple, we choose to work with $k=N$. In this case the chiral squared-mass 
matrices are 
\be
\label{ML2}
\frac{[{\cal{M}}^2_{L}]}{V^2}= 
\left( \begin{array}{cccccccccc}
1 & -\rho^{-1} & 1 & 0 &  & & & & &  \\
-\rho^{-1} & 1+\rho^{-2} 
& -2\rho^{-1} & 1 &  & & & & &  \\
1 & -2\rho^{-1} & 2+ \rho^{-2} 
& -2\rho^{-1}  &  & & & & &\\
0 & 1 & -2\rho^{-1} &  2+\rho^{-2} 
&  & & & & &\\
&&&&\ddots&&&&& \\
&&&&&& 
-2\rho^{-1} &  2+\rho^{-2}   
& -2\rho^{-1}  & 1 \\
&&&&&& 
1 & -2\rho^{-1} & 2+\rho^{-2} & 1-2\rho^{-1}  
\\
&&&&&& 
0 & 1 &  1-2 \rho^{-1} &  2 - 2 \rho^{-1}+  \rho^{-2} 
\end{array} \right) 
\ee
\be
\label{MR2}
\frac{[{\cal{M}}^2_{R}]}{V^2}= 
\left( \begin{array}{cccccccccc}
2+\rho^{-2} & -2\rho^{-1} & 1 &0&  &  & & & &   \\
-2\rho^{-1} & 2+\rho^{-2} 
& -2\rho^{-1}&1  &  & & & & &  \\
1 & -2\rho^{-1} & 2+ \rho^{-2} &-2\rho^{-1}  &  & & & & &\\
&&&&\ddots&&&&& \\
&&&&&&-2\rho^{-1} & 2+\rho^{-2}   
& -2\rho^{-1}  & 1 \\
&&&&&&1 & -2\rho^{-1} & 2+\rho^{-2} & 1-2\rho^{-1}  
\\
&&&&&&0& 1 &  1-2 \rho^{-1} &  2 - 2 \rho^{-1}+  \rho^{-2} 
\end{array} \right) 
\ee
In the analog spring-ball system, $[{\cal{M}}^2_{L,R}]$ represent 
the characteristic matrix of the system's oscillation. The 
mass eigenvalues and mass eigenstates correspond to the proper frequencies 
and the displacement amplitudes of the oscillation. 

\begin{figure}
        \begin{center}    
        \epsfig{figure=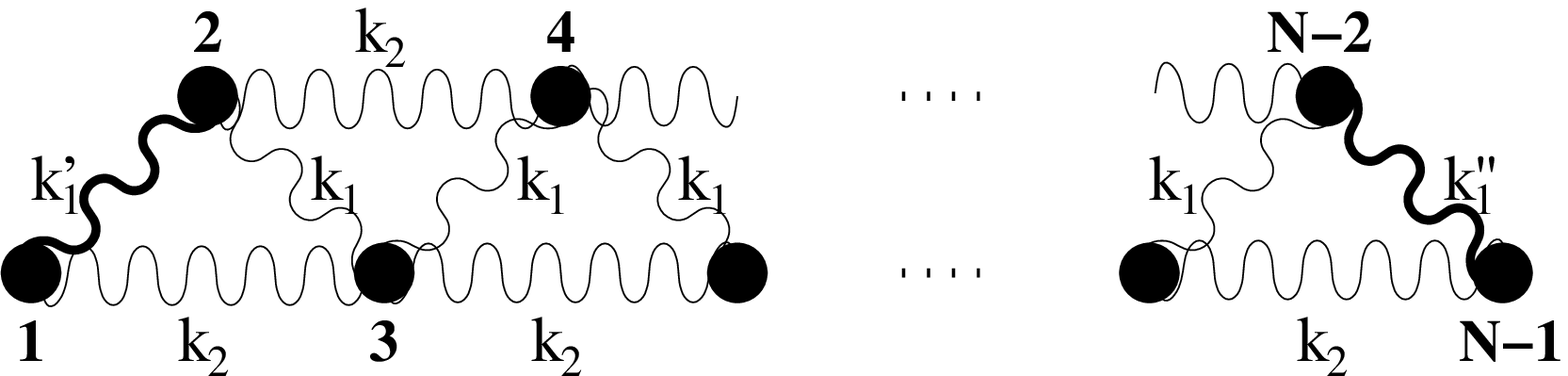,width=0.6\textwidth}
        \end{center}
        \vspace{0cm}
        \caption{Free-end and next-to-nearest-neighbor coupled 
oscillators' system.}
        \label{Fig6}
        \end{figure}

\begin{figure}
        \begin{center}    
        \epsfig{figure=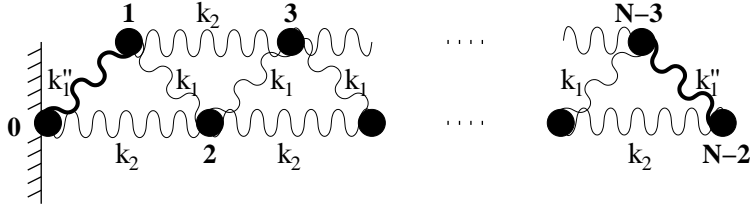,width=0.6\textwidth}
        \end{center}
        \vspace{0cm}
        \caption{One-fixed-end and next-to-nearest-neighbor 
coupled oscillators' system.}
        \label{Fig7}
        \end{figure}

The fact that, for $\rho=1/2$, the sum of all entries in any row of 
$[{\cal{M}}^2_{L}]$, 
and in any row except the first two of $[{\cal{M}}^2_{R}]$, vanishes 
allows us to identify the left sector with a free-end spring-ball system, 
and the right sector with an one-fixed-end system. Indeed,   
the characteristic matrix describing the oscillation of free-end system 
depicted in Fig. \ref{Fig6} explicitly is 
\be
\label{SBML2}
[\Omega_{free}]= 
\left( \begin{array}{cccccccccc}
k'_1+k_2 & -k'_1 & -k_2 & 0 &  & & & & &  \\
-k'_1 & k'_1+k_1+k_2 & -k_1 & -k_2 &  & & & & &  \\
-k_2 & -k_1 & 2k_2+2k_1& -k_1  &  & & & & &\\
0 & -k_2 & -k_1 & 2k_2+2k_1 &  & & & & &\\
&&&&\ddots&&&&& \\
&&&&&&-k_1 & 2k_2+2k_1   & -k_1  & -k_2 \\
&&&&&& -k_2 &-k_1 & k_2+k_1+k_1'' &  -k_1'' \\
&&&&&& 0 & -k_2 &-k_1''  & k_2+k_1''  
\end{array} \right) 
\ee
and the matrix describing the oscillation of one-fixed-end system 
depicted in Fig. \ref{Fig7} 
\be
\label{SBMR2}
[\Omega_{fixed}]= 
\left( \begin{array}{cccccccccc}
k_1''+k_1+k_2 & -k_1 & -k_2 &0&  &  & & & &   \\
-k_1 & 2k_2+2k_1& -k_1  &-k_2 &  & & & & &  \\
-k_2 & -k_1 & 2k_1 +2k_2  &-k_1 & & & & & &\\
&&&&\ddots&&&&& \\
&&&&&&-k_1 & 2k_2+2k_1 & -k_1   & -k_2  \\
&&&&&&-k_2 & -k_1 & k_2+k_1+k_1''  & -k_1'' 
\\
&&&&&&0& -k_2 & -k_1''  &   k_2+k_1''
\end{array} \right) 
\ee
When $\rho=1/2$, we can evidently identify (\ref{ML2}) with (\ref{SBML2}), 
and (\ref{MR2}) with (\ref{SBMR2}) by specifying the springs' constants as 
follows
\be
k_1=\frac{2}{\rho}=4; \;\;\;\;  k_2=-1; \;\;\;\; 
k'_1=\frac{1}{\rho}=2; \;\;\;\; k''_1= -1 + \frac{2}{\rho}=3
\ee
Next we note that, by giving an uniform displacement to the balls, 
the free-end system moves 
from one equilibrium position to another. And since the equilibrium can 
be seen as the oscillation with null frequency, the left squared-mass 
matrix $[{\cal{M}}^2_{L}]$ certainly possesses a zero mass eigenstate and 
a corresponding uniform eigenstate. That is why we have a flat left-handed 
zero mode wave function. In contrast, the fixed-end system does not have 
a translational symmetry, thus $[{\cal{M}}^2_{R}]$ generally does not have 
a zero eigenvalue, i.e. all right-handed fields are massive. For the 
gauge boson sector, and in the Higgsless scenario, the similar 
analog spring-ball systems have been used in \cite{G} to illustrate the 
mass structure of neutral and charged boson towers. 

For any other values of $\rho$, one can multiply each column of 
$[{\cal{M}}^2_{L}]$ by a factor so that the sum of all entries in any row 
of the new formed matrix vanishes. Such factors can always be found 
because the determinant of $[{\cal{M}}^2_{L}]$ is zero as shown in the 
next section. Then we can construct explicitly the new analog  
free-end classical oscillator system, which necessarily possesses a 
zero mode. However, in this general case the ultimate zero mode wave 
function is no longer flat, because its components are obtained by scaling 
back from a uniform profile with different factors. 
The determinant of $[{\cal{M}}^2_{R}]$ does not vanishes thus it cannot be 
attributed to any free-end system, and the right-handed sector does not have 
a massless mode in general. 

In 4D deconstruction, as explained in the appendix, due to the legitimate 
presence of all chiral degrees of freedom, the squared-mass matrices 
$[{\cal{M}}_L^2]$, $[{\cal{M}}_R^2]$ usually correspond to a next-to-nearest 
neighbor interaction of coupled oscillators. As a result, the mass 
spectrum and wave functions of massive modes in general cannot be found 
exactly, although their pertubative expansions can always be obtained. 
Fortunately, there exist methods to identify classes of CBC, whereby the 4D 
deconstruction can be brought down to nearest-neighbor interaction in the 
language of an analog classical system as we shall see next. 
\section{Modeling of massive fermionic states}
In the previous section we have studied the fermion chiral zero mode 
resulting from the imposition of chiral boundary condition of the type 
(\ref{CBC}). Intuitively, because of CBCs (\ref{CBC}), there is one 
more left-handed independent component than the right-handed partners. 
The extra left-handed field cannot be put into a Dirac mass term, 
so it constitutes a chiral zero mode. We now perform a deeper analysis 
on the chiral nature of zero mode, using the linear algebra theorem of 
singular value decomposition (SVD). It turns out that, the theorem also 
leads to the specification of classes of boundary conditions, with which 
the entire mass eigenstate and eigenvalue system can be $exactly$ solved.  
\subsection{Singular Value Decomposition}  
First we present the statement of SVD theorem. Let $[{\cal{S}}]$ be a  
complex matrix of dimension $m \times n$. Without the lost of 
generality we assume $m>n$. Then $[{\cal{S}}]$ can always be written in 
the SVD form: 
$[{\cal{S}}]=[{\cal{U}}][\Sigma][{\cal{V}}]^{\dagger}$, where 
$[{\cal{U}}]$ and $[{\cal{V}}]$ are unitary matrices of dimension 
$m\times m$ and $n\times n$ respectively, and $[\Sigma]$ is a 
$m\times n$-dimension diagonal real matrix 
(i.e. $[\Sigma]_{p,q} \sim \delta_{p,q}$). The proof of this theorem can 
be found e.g. in \cite{GL}.

The SVD theorem allows the chiral mass matrix 
$[{\cal{M}}_{LR}]_{(N-1)\times (N-2)}$ (\ref{mLR}) to be decomposed as
\be
\label{mLR2}
[{\cal{M}}_{LR}]= [{\cal{U}}][\Sigma][{\cal{V}}]^{\dagger}=
[{\cal{U}}]
\left( \begin{array}{cccc}
m_1&&& \\
&&\ddots&\\
&&&m_{N-2} \\
&&&0
\end{array}\right)
[{\cal{V}}]^{\dagger}
\ee
where $\{m_n\}$ is some set of real numbers. Since $[{\cal{M}}_{RL}]= 
[{\cal{M}}_{LR}]^{\dagger}$, we obtain immediately the squared-mass matrices 
defined below Eq. (\ref{DiracEq})
\be
\label{M2L}
[{\cal{M}}^2_{L}]= 
[{\cal{U}}]([\Sigma][\Sigma]^T )[{\cal{U}}]^{\dagger}=
[{\cal{U}}]\left( \begin{array}{cccc}
m_1^2&&& \\
&\ddots&&\\
&&m_{N-2}^2& \\
&&&0
\end{array}\right)[{\cal{U}}]^{\dagger} 
\ee
\be
\label{M2R}
[{\cal{M}}^2_{R}]= 
[{\cal{V}}]([\Sigma]^T[\Sigma] )[{\cal{V}}]^{\dagger}=
[{\cal{V}}]\left( \begin{array}{cccc}
m_1^2&&& \\
&&\ddots&\\
&&&m_{N-2}^2 
\end{array}\right)[{\cal{V}}]^{\dagger}
\ee
Because a unitary rotation leaves the eigenvalues of a matrix unchanged, 
we see clearly that there is a massless left-handed mode, and all massive 
modes come in pair of opposite chiralities. We also see that ${\cal{U}}$ 
and ${\cal{V}}$ actually diagonalizes ${\cal{M}}^2_{L}$ 
and ${\cal{M}}^2_{R}$ respectively.  
\subsection{The construction of fermion higher modes}
The Dirichlet and Neumann boundary conditions (\ref{CBC}) produce a simple, 
in part localized, chiral zero mode of fermion in the framework of 4D 
deconstruction. As in non-universal extra dimension (i.e. brane) models, 
one needs to quantitatively make sure that the contribution of higher 
Kaluza-Klein states to the precisely measured electroweak observables 
is sufficiently small. However, the higher mode structure following these 
CBCs cannot be exactly determined as mentioned earlier. We now attempt to 
specify other classes of CBC, which give rise to an exact and simple 
structure of fermion massive tower, apart form a chiral massless mode. 

First, we begin our construction with the original 
mass matrix $[{\cal{M}}]$ (\ref{MLR}), which 
truly characterizes the mixing of fermion flavors in a vector-like set-up, 
when no boundary 
conditions are imposed. Because it is hermitian, ${\cal{M}}$ can be 
diagonalized by a unitary matrix $[{\cal{U}}_M]$ 
\be
\label{UM}
[{\cal{M}}]=[{\cal{U}}_M]^{\dagger}[{\cal{M}}_D] [{\cal{U}}_M]=
[{\cal{U}}_M]^{\dagger}
\left( \begin{array}{ccc}
M_1&&\\
&\ddots&\\
&&M_N
\end{array}\right)
[{\cal{U}}_M]
\ee
Since $[{\cal{M}}]$ also describes the spring-ball system with only 
nearest-neighbor interaction, $[{\cal{M}}_D]$ and $[{\cal{U}}_M]$ can be 
worked out exactly. Indeed, the eigen-system  
equation associated with $[{\cal{M}}]$ (\ref{MLR}) reads
\be
\label{eqChin}
-V^* U^M_{n-1,k} + M U^M_{n,k} - V U^M_{n+1,k}  = M_k U^M_{n,k}
\;\;\;\;\; (n=0,\ldots,N+1)
\ee
where $M_k$ and $\{U^M_{n,k}\}$ denote respectively the $k$-th eigenvalue 
and eigenstate in the notation of Eq. (\ref{UM}). To streamline the 
presentation we have also introduced extra 
components $U^M_{0,k}$ and $U^M_{N+1,k}$, which are identically zero:  
$U^M_{0,k}=U^M_{N+1,k}=0$. The same method that solved  
Eq. (\ref{eqxn}) now yields  
\beqa
\label{Vm}
M_k &=& M\left(1-2\frac{|V|}{M}\cos\frac{k\pi}{N+1}\right) 
\;\;\;\;\;\;\; (k=1,\ldots,N) \\
\label{Vu}
U^M_{n,k}&=& \sqrt{\frac{2}{N}} e^{-in\theta} \sin\frac{n k\pi}{N+1}
\;\;\;\;\;\;\;\;\;\;\; (n=1,\ldots,N)
\eeqa
Here, we in particular note that when $|\rho|=\frac{|V|}{M} > \frac{1}{2}$, 
there may be an accidental degeneration in the zero mode of the construction 
presented below (i.e. when $M_k=0$ in Eq. (\ref{Vm})). This in turn 
would spoil the single chirality desired for the zero mode under 
construction. It is then natural to assume 
$|\rho| \leq \frac{1}{2}$-an explicit bound on 
$\rho$ that could not have been set by otherwise just looking at the 
localization pattern of the zero mode itself.

Next, from Eqs. (\ref{mLR}), (\ref{mLR2}) and (\ref{UM}) we have the 
following equality
\be
\label{construction}
[{\cal{U}}][\Sigma][{\cal{V}}]^{\dagger}=
[{\cal{B}}_L]^{\dagger}[{\cal{U}}_M]^{\dagger}[{\cal{M}}_D] [{\cal{U}}_M]
[{\cal{B}}_R]
\ee
On the left-hand side, $[\Sigma]$ as defined in Eq. (\ref{mLR2}) contains 
all $N-2$ eigen-masses $\{m_n\}_1^{N-2}$ of massive modes that we want to determine. 
On the right-hand side, $[{\cal{M}}_D]$ (\ref{UM}) contains $N$ known 
eigen-masses (\ref{Vm}) of the vector-like model. In a straightforward construction, we 
wish to identify $\{m_n\}_1^{N-2}$ with $N-2$ eigenvalues of 
vector-like matrix 
$[{\cal{M}}]$, say {$\{M_n\}_1^{N-2}$}. This can be realized in many 
different ways. Just for the purpose of illustration, we present below 
a particular, not necessarily simplest, choice of SVD unitary matrices 
$[{\cal{U}}]$, $[{\cal{V}}]$ and the boundary conditions $[{\cal{B}}_L]$, 
$[{\cal{B}}_R]$ that fulfill this identification requirement.
\be
\label{VU}
[{\cal{V}}]= {\bf{1}}_{(N-2)\times(N-2)}; \;\;\;\;\;
[{\cal{U}}]=
\left(
\begin{array}{ccc}
{\bf{1}}_{(N-3)\times(N-3)}&&\\
&\frac{1}{\sqrt{2}}&\frac{1}{\sqrt{2}}\\
&-\frac{1}{\sqrt{2}}&\frac{1}{\sqrt{2}}
\end{array}\right)
\ee
\be
\label{BL}
[{\cal{B}}_L]_{N\times(N-1)}=\left(\begin{array}{cccc}
U^M_{1,1} &\ldots&\ldots& U^M_{1,N}\\
:&\ldots&\ldots&:\\
U^M_{N-3,1} &\ldots&\ldots& U^M_{N-3,N} \\
U^M_{N-2,1}/\sqrt{2} &\ldots&\ldots& U^M_{N-2,N}/\sqrt{2} \\
-U^M_{N-2,1}/\sqrt{2} &\ldots&\ldots& -U^M_{N-2,N}/\sqrt{2}
\end{array}\right)^{\dagger}
\ee
\be
\label{BR}
[{\cal{B}}_R]_{N\times(N-2)}=\left(\begin{array}{cccc}
U^M_{1,1} &\ldots&\ldots& U^M_{1,N}\\
:&\ldots&\ldots&:\\
U^M_{N-3,1} &\ldots&\ldots& U^M_{N-3,N} \\
U^M_{N-2,1} &\ldots&\ldots& U^M_{N-2,N}
\end{array}\right)^{\dagger}
\ee
where $U^M_{k,q}$ is the element in the $k$-th row and $q$-th column of 
the matrix $[{\cal{U}}_M]$ (Eq. (\ref{Vu})) that diagonalizes the 
matrix ${\cal{M}}$ (\ref{UM}). Because $[{\cal{U}}_M]$ is unitary, we 
can verify that Eq. (\ref{construction}) holds for the choice 
(\ref{VU})-(\ref{BR}). 

As a result of this particular construction, the spectrum of 
deconstructed fermions contains one left-handed zero mode and 
$N-2$ higher vector-like modes of the masses identical to the 
eigenvalues $M_k$ ($k=1,\ldots ,N-2$) (Eq. (\ref{Vm})) of the original 
vector-like mass matrix $[{\cal{M}}]$. 

The combination of Eqs. (\ref{CBCL}), 
(\ref{CBCR}), (\ref{Vu}), (\ref{BL}) and (\ref{BR})  gives  
the explicit CBC needed in the realization of this construction 
\footnote{Any boundary condition can be written in a gauge-invariant form 
by incorporating an appropriate product of link fields, e.g. CBC 
$\frac{\phi_{k,k+1}}{V}\ldots \frac{\phi_{q-1,q}}{V} \psi_{qL}= \psi_{kL}$ 
is gauge-invariant and implies a simple relation $\psi_{qL}= \psi_{kL}$ 
after the spontaneous symmetry breaking $\langle\phi\rangle = V$}. Generally, 
from Eqs. (\ref{M2L}), (\ref{M2R}) and the discussion below 
Eq. (\ref{DiracEq}) we see that the corresponding wave functions of left and 
right-handed eigen-modes are just the column vectors of SVD unitary matrices 
${\cal{U}}$ and ${\cal{V}}$ respectively. Specifically for the choice 
(\ref{VU})-(\ref{BR}), most of states in mass eigenbasis are identical to 
those in flavor eigenbasis as indicated by the explicit expression 
(\ref{VU}). The only two exceptions are the massless and $(N-2)$-th massive  
left-handed states, which are written in flavor space as (see Eqs. 
(\ref{WF}), (\ref{VU}))
\be
\label{2WF}
\tilde{\psi}_{0L} = \frac{1}{\sqrt{2}}( \psi_{N-2L} +  \psi_{N-1L} );
\;\;\;\;\;\;\;
\tilde{\psi}_{N-2L} = \frac{1}{\sqrt{2}}( \psi_{N-2L} -  \psi_{N-1L} ) 
\ee
As before, $\tilde{\psi}$ and $\psi$ respectively denote mass and flavor 
eigenstates. 

The mass eigenstates (\ref{2WF}) may look rather simple, but we can 
easily and 
systematically make all mass eigenstates more involved into the flavor 
space as follows. First we introduce the new boundary conditions 
$[{\cal{B}}_L] \rightarrow [{\cal{B}}'_L] 
\equiv [{\cal{B}}_L][{\cal{U}}_{L}] $; 
$[{\cal{B}}_R] \rightarrow [{\cal{B}}'_R] 
\equiv [{\cal{B}}_R][{\cal{U}}_{R}] $ 
where $[{\cal{U}}_{L}]$ and $[{\cal{U}}_{R}]$ are some 
unitary matrices of dimension $(N-1)\times (N-1)$ and 
$(N-2)\times (N-2)$ respectively. The chiral squared-mass matrices 
accordingly transform as (see Eq. (\ref{mLR})) $[{\cal{M}}_L^2] \rightarrow 
[{{\cal{M}}'}_L^2] = 
[{\cal{U}}_{L}]^{\dagger}[{\cal{M}}_L^2] [{\cal{U}}_{L}]$ 
and $[{\cal{M}}_R^2] \rightarrow [{{\cal{M}}'}_R^2] = 
[{\cal{U}}_{R}]^{\dagger}[{\cal{M}}_R^2] [{\cal{U}}_{R}]$. Obviously, 
this unitary transformation does not alter the mass structure of 
fermions. New and old CBC then can be referred to as being in the same 
class of {\em unitarily equivalent} boundary conditions. In contrast, this  
change of boundary condition deeply affects the distribution of mass 
eigenstates in flavor space. For the old boundary condition it is 
$\{\tilde{\psi}_L\}= [{\cal{U}}]^{\dagger}\{\psi_L\}$ and 
$\{\tilde{\psi}_R\}= [{\cal{V}}]^{\dagger}\{\psi_R\}$ respectively for 
left and right sectors. For the new boundary condition, it becomes 
$\{\tilde{\psi}'_L\}= [{\cal{U}}]^{\dagger}[{\cal{U}}_L]\{\psi_L\}$ and 
$\{\tilde{\psi}'_R\}= [{\cal{V}}]^{\dagger}[{\cal{U}}_R]\{\psi_R\}$. Making 
appropriate choice of matrices $[{\cal{U}}_{L}] \in U(N-1)$ and 
$[{\cal{U}}_{R}] \in U(N-2)$, we can flexibly modify the appearance of 
all mass eigenstates in the flavor space, but without changing the mass 
eigenvalues themselves. Because the 4D propagation of  
massive particles is proportional to the inverse of their squared mass, the   
suppression of exotic fermion higher mode contribution primarily 
depends on their masses. Hence  
the construction above can have relevant applications in model building.  
Through the boundary conditions, we can alter the shape of chiral 
zero mode while still keeping under a permanent suppression all the 
effects induced by higher modes of a prefixed mass spectrum. In 
the next section we will illustrate this point in the building of quark mass 
hierarchy.

\subsection{Application: Complex CKM mixing via chiral boundary condition}
In this section we construct the mass hierarchy of SM quarks by having 
different ``overlaps'' of zero-mode fermion in flavor space. It would be 
interesting to apply the results derived in Section II to the construction 
of a model of fermion masses, using the idea of ``localization'' in group 
index space. This has been carried out in Ref. \cite{HST} 
(see also \cite{AKMY}). In the current work, we adopt a different approach 
to emphasize the effect of deconstruction boundary conditions. 
Here all quarks involved come from a single class 
of unitarily equivalent boundary conditions, thus they have the same mass 
spectrum as in the universal ED scenario. The distinct wave functions, 
and hence their overlaps as well as the CP violating phase, all are encoded 
concisely in the boundary conditions. Another mechanism to 
generate fermion mass hierarchy using arbitrary link field transformations 
has been proposed in \cite{NOS}.

We first identify the deconstruction product group as 
$\prod_{n=1}^{N} G_n \equiv \prod_{n=1}^{N} [SU(2)\times U(1)]_n$. 
To each member $[SU(2)\times U(1)]_n$ of the group chain we associate a 
$SU(2)_n$-doublet scalar Higgs $H_n$, three $SU(2)_n$-doublets $Q^{(i)}_n$ and 
six $SU(2)_n$-singlets $U^{(i)}_n$, 
$D^{(i)}_n$ of spin $\frac{1}{2}$, just like in the standard model 
(with family index $i=1,2,3$). The deconstruction interaction 
among $Q^{(i)}_n$ themselves (or $U^{(i)}_n$, or $D^{(i)}_n$) and 
the appropriate CBCs give rise to a zero mode of 
desired chirality as required by SM. The Yukawa interaction with Higgs 
fields next generates the mass for these zero modes. In this scenario, it is 
tempting to place the source of mass hierarchy solely in the difference of 
zero-mode distribution in DD group index space, 
so we assume the universal Yukawa couplings within Up and Down sectors. From 
the gauge-invariant Yukawa terms
\be
\label{Ygauge}
\kappa_U \sum_{n=1}^{N} \bar{Q}^{(i)}_n i\sigma_2 H_n^{*} U^{(j)}_n +
\kappa_D \sum_{n=1}^{N} \bar{Q}^{(i)}_n  H_n D^{(j)}_n + H.c.
\ee 
and given a uniform VEV of all Higgses $\langle H_n\rangle = h$ in 
accordance with the permutation symmetry, we can extract the mass terms 
for zero modes
\be
h\kappa_U\sum_{i,j=1}^{3} \bar{\tilde{Q}}^{(i)}_0 M^U_{ij} \tilde{U}^{(j)}_0
+h\kappa_D\sum_{i,k=1}^{3} \bar{\tilde{Q}}^{(i)}_0 M^D_{ik} \tilde{D}^{(k)}_0
\ee
where
\beqa
\label{Mu}
M^U_{ij} = \sum_{n=1}^{N-1} [{\cal{U}}^{(i)}_Q]^*_{nN-1}
[{\cal{U}}^{(j)}_U]_{nN-1}= 
[{\cal{U}}^{(i)\dagger}_Q {\cal{U}}^{(j)}_U]_{N-1,N-1}  \\ 
\label{Md}
M^D_{ij} = \sum_{n=1}^{N-1} [{\cal{U}}^{(i)}_Q]^*_{nN-1}
[{\cal{U}}^{(j)}_D]_{nN-1}= 
[{\cal{U}}^{(i)\dagger}_Q {\cal{U}}^{(j)}_D]_{N-1,N-1}
\eeqa
To arrive to these expressions for zero-mode mass matrices we have used 
the facts that ${\cal{U}}^{(i)}_Q$ 
diagonalizes the squared-mass matrix $[{\cal{M}}^{2(i)}_{QL}]$ (\ref{M2L}), 
and that in the convention of (\ref{M2L}), the 
zero eigenvalue is placed in the bottom right entry of a 
matrix of dimension $(N-1)\times (N-1)$. Similar notations 
for $U$ and $D$ fields have been also assumed. 

As discussed in the last part of previous section, the zero-mode coefficients 
$[{\cal{U}}^{(j)}_{Q,U,D}]_{nN-1}$ can be easily regulated 
by modifying the boundary condition. 
We consider here a simple three-site gauge model $[SU(2)\times U(1)]^3$ 
(i.e. $N=3$). The CBCs on all quarks in general are different, we however 
assume that they belong to a single unitarily equivalent class 
(see Section IIIB), which also includes the ``reference'' 
Dirichlet-Neumann CBC (\ref{CBC}). In this case, when $\rho =1/2$, 
the actual wave function of any zero-mode quark is recovered 
from the ``reference'' zero mode by a rotation  
\footnote{The rotation is completely characterized by the CBC on the 
corresponding quark, as shown in Section IIIB.} 
\be
\label{Q}
\left\{[{\cal{U}}^{(i)}_{Q}]_{n,N-1}\right\}_{n=1}^{N-1} = 
\exp{\left( i \vec{\alpha}_Q^{(i)} \frac{\vec{\tau}}{2}\right)}  
\left\{ \frac{1}{\sqrt{N-1}} \right\}_1^{N-1}
\ee
where $N=3$ and $\tau$'s denote Pauli matrices. The uniform 
$(N-1)$-dimensional vector 
$\left\{ \frac{1}{\sqrt{N-1}} \right\}_1^{N-1}$ represents the  
``reference'' flat zero mode (\ref{solxn}) obtained in Section II 
for the Dirichlet-Neumann CBC 
\footnote{Within the three-site gauge model $[SU(2)\times U(1)]^3$ 
under consideration, there is no interaction that can violate the baryon 
number. Moreover, even if a larger GUT scenario is incorporated at 
some higher energy scale, one can always choose appropriate CBCs and 
parameter $\rho$ for lepton sector to produce a sufficiently small 
quark-lepton mixing interaction. Since a more quantitative treatment of lepton 
sector lies beyond the scope of this paper, here we will not pursue 
further the issues of baryogenesis and proton decay.}. 
The generic $SU(N-1)$ matrix characterized by 
$\vec{\alpha}=(\alpha_x,\alpha_y,\alpha_z)$ implements the deviation of the 
actual CBC from that of the reference configuration for the case $N=3$. 
The characterization similar to (\ref{Q}) can also be given for $U$ and $D$ 
fields. Plugging (\ref{Q}) into (\ref{Mu}), (\ref{Md}) we obtain 
the mass matrix elements
\beqa
\label{mu}
M^U_{ij}=\frac{1}{N-1}\sum_{n,k=1}^{3} 
\left[\exp{\left( i (-\vec{\alpha}_Q^{(i)}+\vec{\alpha}_U^{(j)} )
\frac{\vec{\tau}}{2}\right)} \right]_{n,k} \\
\label{md}
M^D_{ij}=\frac{1}{N-1}\sum_{n,k=1}^{3} 
\left[\exp{\left( i (-\vec{\alpha}_Q^{(i)}+\vec{\alpha}_D^{(j)} )
\frac{\vec{\tau}}{2}\right)} \right]_{n,k}
\eeqa
Using the equality 
$\exp{(i \vec{\alpha} \frac{\vec{\tau}}{2})} = 
\cos{(|\alpha|/2)} + 
i \frac{\vec{\alpha}}{|\alpha|}\vec{\tau} \sin{(|\alpha|/2)} $
we can write (\ref{mu}), (\ref{md}) more explicitly as
\beqa
\label{emu}
M^U_{ij}=\cos{\frac{|-\alpha^{(i)}_Q+\alpha^{(j)}_U|}{2}}+
i\;\frac{-\alpha^{(i)}_{Qx}+\alpha^{(j)}_{Ux}}
{|-\alpha^{(i)}_Q+\alpha^{(j)}_U|}
\sin{\frac{|-\alpha^{(i)}_Q+\alpha^{(j)}_U|}{2}} \\
\label{emd}
M^D_{ij}=\cos{\frac{|-\alpha^{(i)}_Q+\alpha^{(j)}_D|}{2}}+
i\;\frac{-\alpha^{(i)}_{Qx}+\alpha^{(j)}_{Dx}}
{|-\alpha^{(i)}_Q+\alpha^{(j)}_D|}
\sin{\frac{|-\alpha^{(i)}_Q+\alpha^{(j)}_D|}{2}}
\eeqa
where $|\alpha| \equiv \sqrt{\alpha_x^2+\alpha_y^2+\alpha_z^2}$ and $N=3$ 
have been employed. 

First, the real and imaginary parts of each mass element 
are well independent, because once $|-\alpha^{(i)}_Q+\alpha^{(j)}_U|$ is 
fixed, we still have the freedom to choose 
$(-\alpha^{(i)}_{Qx}+\alpha^{(j)}_{Ux})/|-\alpha^{(i)}_Q+\alpha^{(j)}_U|$ 
to be anywhere between $-1$ and $1$. That is each of $M^U_{ij}$ 
(and $M^D_{ij}$) can practically have an arbitrary complex phase. Once we 
diagonalize the matrices $[M^U M^{U\dagger}]$ and $[M^D M^{D\dagger}]$, those 
phases will combine and transform themselves into the phases of CKM matrix 
to give rise to the CP violation processes via the well-known 
Kobayashi-Maskawa mechanism. Second, we also see that (\ref{emu}), 
(\ref{emd}) can accommodate a vast class of mass matrix hierarchy. 
Just to mention some simplest examples, when 
$\vec{\alpha}^{(i)}_Q= \vec{\alpha}^{(j)}_U$ we obtain a democratic 
mass matrix (see e.g. \cite{KM}), and when $\vec{\alpha}_Q$ and 
$\vec{\alpha}_U$ point along the $x$-direction 
($|\alpha^{(i)}_{Q}|= \alpha^{(i)}_{Qx}$, 
$|\alpha^{(j)}_{U}|= \alpha^{(j)}_{Ux}$), a pure phase mass matrix 
\cite{BSR,FH,HSS} for up sector. 
Numerical analysis on the realization of fermion mass hierarchy 
via 4D deconstruction is presented in details in a parallel work \cite{HST}.  
\section{Conclusion}
We have investigated the structure of fermion states in their mass 
eigenbasis following a pure 4-dimensional deconstruction process. The zero 
modes expose some interesting localization pattern, that can have useful 
applications in phenomenology. It is found that the ratio of the two 
underlying UV breaking scales in the deconstruction scenario is primarily 
responsible for such localization. The chiral boundary conditions used to 
obtain chiral zero mode also have crucial influence on the structure of the 
whole spectrum. CBCs can be put in different classes, each of which 
characterizes a single mass spectrum. The modification in CBC leads to the 
change in wave functions and their overlaps in the flavor space. This 
interactive relation can be used to model the complex CKM mixing of 
fermion flavors in the Standard Model.

In this paper we have identified the importance of CBC, but have not 
discussed their generation mechanism. We hope to come back to this   
as well as to other related issues concerning gauge boson sector  
in 4D deconstruction in future work.   
\begin{acknowledgments}
This work is supported in part by the U.S. Department of Energy under Grant 
No. DE-A505-89ER40518. N-K.T. also acknowledges the Dissertation Year 
Fellowship from UVA Graduate School of Arts and Sciences.
\end{acknowledgments}
\begin{appendix}
\section{Latticized Extra Dimension and Deconstruction}
We consider in this appendix a simple deconstruction scenario with 
a product of Abelian groups 
$\prod_{n=1}^N U(1)_n$. The argument can be generalized to non-Abelian case. 
The connection of this scenario to a latticized extra dimension is based on 
the observation that the deconstruction link field and the Wilson line along 
extra dimension have similar gauge transformation. Therefore we 
perform the following identification  
\be
\phi_{n,n+1} \sim e^{ig\int_{na}^{(n+1)a}dy A_5 }
\ee
where $g$ denotes the gauge coupling, $a$ the lattice spacing, $A_5$ the 
extra dimensional component of gauge boson, $y$ the coordinate of extra 
dimension. The latticization of the fifth dimension also allows us to 
write $\partial_5 \psi \equiv \partial \psi/ \partial y \sim 
(\psi_{n+1}- \psi_{n-1})/2a$, so that in the gauge $A_5=0$ the extra 
dimensional 
piece of kinetic term becomes
\be
\label{d5}
\bar{\psi} \gamma^5 \partial_5 \psi \sim \frac{1}{2a} \left(
\bar{\psi}_n \gamma^5 \phi_{n,n+1} \psi_{n+1} + H.c.\right)
\ee
where the link field has been inserted to assure the gauge invariance. The 
$\gamma_5$ is a remnant of the fifth dimension and leads to a profound 
difference between latticized ED and pure 4D deconstruction as we will see 
below. By using (\ref{d5}), the fermionic mass 
term of the latticized Lagrangian can be written (apart from a bare mass) 
in the chiral basis as
\be
\label{Lmass}
\sum_{n=1}^N \left(\bar{\psi}_{nR} \phi_{n,n+1} {\psi}_{n+1L} -
\bar{\psi}_{nL} \phi_{n,n+1} {\psi}_{n+1R} + H.c.\right) 
\ee 
We again note that the difference in the sign of these two terms originates 
from the chiral matrix $\gamma_5$. The mass spectrum induced 
by (\ref{Lmass}) is found to be (see \cite{HL})
\be
M_n^2 = M^2 + V^2 \sin^2[(2n+1)\pi/N]
\ee
where $V$ is the VEV of link fields and $M$ is the bare mass that has not 
explicitly written in (\ref{Lmass}). 
It is  argued in \cite{HL} that half of these modes results from a lattice 
artifact and stands for the fermion flavor doubling problem of lattice 
gauge theory. To remove the spurious fermionic modes, it is proposed to add 
the Wilson term $\eta\bar{\psi} (\partial_5 +igA_5)^2 \psi /V $ to the 
5D Lagrangian, where $\eta$ is some dimensionless constant. 
In the result, (\ref{Lmass}) now becomes
\be
\label{WLmass}
\sum_{n=1}^N \left[(1-\eta)\bar{\psi}_{nR} \phi_{n,n+1} {\psi}_{n+1L} -
(1+\eta)\bar{\psi}_{nL} \phi_{n,n+1} {\psi}_{n+1R} + H.c. \right] 
\ee 
Choosing $\eta=1$ or $-1$ one can solve the flavor doubling problem but 
obviously either choices also eliminate half of fermionic chiral degree of 
freedom. Therefore in the models inspired by latticized extra 
dimension, fermions to begin 
with can be chosen to be Weyl spinors. In the 4D deconstruction approach, 
however, 
the whole spectrum of fermions is physical and we retain twice as many 
chiral modes. That is, fermions to start with are all Dirac spinors.
\end{appendix}

\end{document}